\documentclass[twoside,leqno,twocolumn]{article}

\usepackage[letterpaper]{geometry}

\usepackage[numbers,sort]{natbib}
\usepackage{ltexpprt}
\usepackage{hyperref}

\usepackage{booktabs}   
\usepackage{subcaption} 

\usepackage{latexsym,amsmath,textcomp,pifont,marvosym,wasysym,amssymb}
\usepackage{graphicx}

\usepackage{comment}

\usepackage{tikz}
\usetikzlibrary{calc}

\usepackage{pgfplots}
\usepgfplotslibrary{external}
\tikzsetexternalprefix{plots/}

\newif\ifpdfplots
\pdfplotstrue

\usepackage[algo2e,ruled,vlined,noend]{algorithm2e}
\SetCommentSty{textit}
\DontPrintSemicolon
\IncMargin{-\parindent}
\SetAlCapHSkip{0pt}
\SetAlgoLined

\SetKwFor{ParallelFor}{for}{do in parallel}{}
\SetKwIF{If}{ElseIf}{Else}{if}{}{else if}{else}{end if}

\usepackage{xcolor}
\usepackage{pifont}

\newcommand{\Oh}[1]{\ensuremath{\mathcal{O}(#1)}}

\newcommand{\Partition}{\ensuremath{\mathrm{\Pi}}}%
\newcommand{\incnets}{\ensuremath{\mathrm{I}}}%

\newcommand{\con}{\ensuremath{\lambda}}
\newcommand{\conset}{\ensuremath{\Lambda}}

\newcommand{\pluseq}{\mathrel{+}=}
\newcommand{\minuseq}{\mathrel{-}=}
\newcommand{\argmax}{\operatorname{arg \, max}}

\definecolor{darkgreen}{rgb}{0.0, 0.6, 0.0}

\newcommand{\kommentar}[1]{}

\newcommand{\nikolai}[1]{\kommentar{\color{darkgreen}[NM: #1]}}

\newcommand{\review}[1]{\kommentar{\color{purple}[Review: #1]}}

\newcommand{\splitatcommas}[1]{%
  \begingroup
  \begingroup\lccode`~=`, \lowercase{\endgroup
    \edef~{\mathchar\the\mathcode`, \penalty0 \noexpand\hspace{0pt plus 1em}}%
  }\mathcode`,="8000 #1%
  \endgroup
}

\begin{document}

	\newcommand\relatedversion{}
	\renewcommand\relatedversion{\thanks{The full version of the paper can be accessed at \protect\url{https://arxiv.org/abs/1902.09310}}} 

	\title{Deterministic Parallel High-Quality Hypergraph Partitioning}
	
	\author{
		Robert Krause\thanks{Karlsruhe Institute of Technology, Karlsruhe, Germany. robert.krause@student.kit.edu, nikolai.maas@kit.edu}  \and 
		Lars Gottesbüren\thanks{Google Research, Zürich. gottesbueren@google.com} \and
		Nikolai Maas\footnotemark[1]
	}
	\date{}
	
	\maketitle
	
	
	\fancyfoot[R]{}
	
	
	
	
	
\begin{abstract} \small\baselineskip=9pt

We present a deterministic parallel multilevel algorithm for balanced hypergraph partitioning that matches the state of the art for non-deterministic algorithms. Deterministic parallel algorithms produce the same result in each invocation, which is crucial for reproducibility. Moreover, determinism is highly desirable in application areas such as VLSI design. While there has been tremendous progress in parallel hypergraph partitioning algorithms recently,
deterministic counterparts for high-quality local search techniques are missing. Consequently, solution quality is severely lacking in comparison to the non-deterministic algorithms.

In this work we close this gap. First, we present a generalization of the recently proposed Jet refinement algorithm. While Jet is naturally amenable to determinism, significant changes are necessary to achieve competitive performance on hypergraphs. We also propose an improved deterministic rebalancing algorithm for Jet. Moreover, we consider the powerful but slower flow-based refinement and introduce a scheme that enables deterministic results while building upon a non-deterministic maximum flow algorithm.

As demonstrated in our thorough experimental evaluation, this results in the first deterministic parallel partitioner that is competitive to the highest quality solvers. With Jet refinement, we match or exceed the quality of Mt-KaHyPar's non-deterministic default configuration while being only 15\% slower on average. We observe self-relative speedups of up to 55 on 64 cores with a 22.5$\times$ average speedup.
Our deterministic flow-based refinement exceeds the quality of the non-deterministic variant by roughly 1\% on average but requires 31\% more running time.

\end{abstract}

\section{Introduction}

\begin{figure}
    \begin{minipage}{\textwidth}
    \ifpdfplots
    \includegraphics{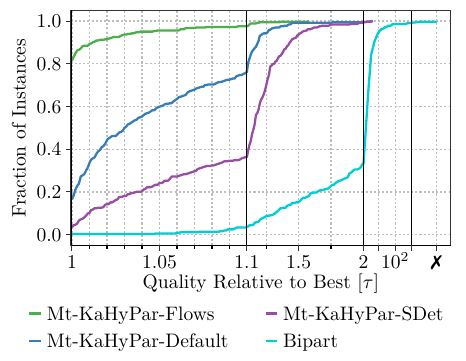}
    \else
    \tikzsetnextfilename{quality_initial}%
    \input{tikz/quality_initial}%
    \fi

    \end{minipage}
    \vspace{-10pt}
    \caption{The quality gap between non-deterministic solvers (Mt-KaHyPar-\{Default, Flows\}) and deterministic solvers (BiPart, Mt-KaHyPar-SDet). Higher and to the left is better. Refer to Section~\ref{sec:exp-setup} for details on the plot.}
    \label{fig:quality_initial}
\end{figure}

Balanced hypergraph partitioning is a fundamental optimization problem with applications in key domains such as VLSI design~\cite{ALPERT-SURVEY}, scientific computing~\cite{PATOH, ZOLTAN}, distributed database sharding~\cite{SOCIALHASH} and more~\cite{HYPERGRAPH-SURVEY}.
The task is to divide the vertices of a hypergraph into roughly equal-sized blocks such that the \emph{connectivity metric} is minimized, which is defined as the sum of the number of different blocks connected by each hyperedge.
As balanced partitioning is NP-hard to approximate within a constant factor~\cite{AndreevRaeckeApprox}, the prevalent technique for obtaining good solutions in practice is the multilevel method~\cite{HENDRICKSON}.
The multilevel scheme first constructs a series of increasingly smaller hypergraphs by contracting vertices in the \emph{coarsening phase}, and then finds a good \emph{initial partition} on the smallest hypergraph.
Subsequently, the contractions are undone in reverse order in the \emph{uncoarsening phase}.
The current partition is projected to the next larger hypergraph and local search algorithms are used to improve the partition.
The improvement step is called \emph{refinement}.

A recent line of work~\cite{mt-kahypar-d, mt-kahypar-q, mt-kahypar-flows, mt-kahypar-journal} developed multi-core parallel algorithms that match the solution quality of the highest quality sequential methods~\cite{kahypar-hfc, KAHYPAR-K}.
Yet, these are still largely non-deterministic.
Deterministic algorithms always produce the same output in each invocation when the same input is used.
Researchers have long advocated the benefits of determinism in parallel algorithms~\cite{DBLP:conf/popl/Steele90, DBLP:journals/computer/Lee06, bocchino2009parallel, DBLP:conf/ppopp/BlellochFGS12} such as reasoning about performance, ease of debugging and reproducibility.
In some applications reproducibility is necessary. 
For example in VLSI design, hardware engineers perform manual post-processing steps which expect a certain initial solution, and which are expensive to repeat~\cite{BIPART}.

Contrary to the sequential setting where using pseudo-random number generation with a fixed random seed suffices, it can be highly non-trivial to make parallel programs deterministic.
In particular, changes to the internal workings of the algorithms may be required, which can lead to performance degradation compared to the non-deterministic counterparts~\cite{DBLP:conf/ppopp/BlellochFGS12}.
Partitioning algorithms work with an existing partition and move vertices between blocks based on \emph{gains} -- the improvement in the objective function.
Traditionally, parallel vertex moves are performed asynchronously for performance and solution quality reasons~\cite{mt-kahypar-d}, leading to race conditions in the computed gains, which in turn lead to non-deterministic partitioning results~\cite{mt-kahypar-sdet}.

There are two existing works on deterministic parallel partitioning: BiPart~\cite{BIPART} and Mt-KaHyPar's deterministic mode~\cite{mt-kahypar-sdet}.
However, they achieve worse solution quality than state-of-the-art methods, as shown in Figure~\ref{fig:quality_initial}.
This is due to the lack of high-quality refinement algorithms such as parallel localized FM~\cite{MT-KAHIP, mt-kahypar-d} and flow-based refinement\cite{KAFFPA, mt-kahypar-flows, kahypar-hfc, REBAHFC}.
 Instead, both deterministic partitioners use a \emph{synchronous label propagation} refinement, which only considers moves with positive gain and is thus unable to escape local minima -- contrary to FM refinement, which builds a sequence of dependent moves that can include negative gain moves.
Flow-based refinement is even more powerful as it takes a more global view of the optimization problem by computing minimum cuts on cut boundary segments. 

\paragraph{Contributions}

In this paper, we close the quality gap by presenting two new deterministic versions of Mt-KaHyPar: DetJet and DetFlows.
In place of FM refinement, we employ the recently proposed Jet refinement~\cite{Jet-Journal}, an algorithm designed for graph partitioning on GPUs with the goal of matching the solution quality of FM.
This is achieved through unconstrained local search, i.e., allowing large temporary imbalances that are then repaired in a rebalancing step.
As a GPU algorithm, it is naturally suited for a deterministic implementation.
Here our technical contributions are an improved and deterministic rebalancing algorithm, as well as an efficient generalization of Jet from graphs to hypergraphs.
Specifically, a naive generalization of Jet's so-called \emph{afterburner} component incurs $O(\sum_{e \in E} |e|^2)$ work, which is infeasibly slow.
We present an implementation with $O(\sum_{e \in E} |e|\log|e|))$ work.

Secondly, we devise a deterministic version of Mt-KaHyPar's flow-based refinement. 
Our contributions are a matching-based scheduling scheme of independent block pairs for two-way refinement and low-level adaptations of the two-way refinement scheme.
The latter permit using a non-deterministic maximum flow algorithm, which is crucial for performance, while guaranteeing deterministic bipartition results.

Finally, we implement two small improvements to Mt-KaHyPar's deterministic coarsening that further improve solution quality and robustness.
Mt-KaHyPar-DetJet uses the improved coarsening and deterministic Jet for refinement.
Mt-KaHyPar-DetFlows additionally uses deterministic flow-based refinement.

\paragraph{Results}

Mt-KaHyPar-DetJet matches or exceeds the solution quality of Mt-KaHyPar-Default at a 15\% increase in running time.
Mt-KaHyPar-DetFlows achieves a 1\% improvement in solution quality over Mt-KaHyPar-Flows at a 31\% increase in running time.
Notably, Mt-KaHyPar-Flows achieves solution quality on the same level as the highest quality sequential solvers~\cite{kahypar-journal, KAFFPA} while being substantially faster through parallelism, e.g., an order of magnitude on 10 cores~\cite{mt-kahypar-flows}. By extension the same applies to our  deterministic parallel variant.
We perform strong scaling experiments for Mt-KaHyPar-DetJet, showing up to 55$\times$ self-relative speedup on 64 cores, with an average speedup of 22.5.
In an individual comparison with BiPart, Mt-KaHyPar-DetJet outperforms BiPart by a factor of 2.4 in solution quality and 3.5 in running time, computing better partitions on $99.6\%$ of the instances.

\section{Preliminaries}\label{sec:preliminaries}

\paragraph{Hypergraphs}

A \emph{weighted hypergraph} $H=(V,E,c,\omega)$ is a set of vertices $V$ and a set of hyperedges $E$ with vertex weights $c:V \to \mathbb{N}$ and hyperedge weights $\omega:E \to \mathbb{N}$, where each hyperedge $e$ is a subset of the vertex set $V$.
The vertices of a hyperedge are called its \emph{pins}.
A vertex $v$ is \emph{incident} to a hyperedge $e$ if $v \in e$.
$\incnets(v)$ denotes the set of all incident hyperedges of $v$.
The \emph{degree} of a vertex $v$ is $d(v) := |\incnets(v)|$.
The \emph{size} $|e|$ of a hyperedge $e$ is the number of its pins.
We extend $c$ and $\omega$ to sets in the natural way $c(U) :=\sum_{v\in U} c(v)$ and $\omega(F) :=\sum_{e \in F} \omega(e)$.

\paragraph{Partitions}

A \emph{$k$-way partition} of a hypergraph $H$ is a function $\Partition : V \to [k]$ that assigns each vertex to a \emph{block} (or block identifier) $i \in [k]$ \review{please define [k], since I wonder whether the set
    is 0-based ({0,1...,k-1}) or 1-based}.
In addition to the identifiers, we also use the term block for their corresponding vertex sets $V_i := \Partition^{-1}(i)$.
We call $\Partition$ \emph{$\varepsilon$-balanced} if each block $V_i$ satisfies the \emph{balance constraint}:
$c(V_i) \leq L_{\max} := (1+\varepsilon) \lceil c(V) / k \rceil$ for some parameter $\varepsilon \in (0,1)$.

For each hyperedge $e$, $\conset(e) := \{V_i \mid  V_i \cap e \neq \emptyset\}$ denotes the \emph{connectivity set} of $e$.
The \emph{connectivity} $\con(e)$ of a hyperedge $e$ is $\con(e) := |\conset(e)|$.
A hyperedge is called a \emph{cut hyperedge} if $\con(e) > 1$.
Given parameters $\varepsilon$, and $k$, and a hypergraph $H$, the \emph{hypergraph partitioning problem} is to find an $\varepsilon$-balanced $k$-way partition $\Partition$ that minimizes the \emph{connectivity metric} $(\lambda - 1)(\Pi) := \sum_{e \in E} (\lambda(e) - 1) \: \omega(e)$.
We use the term solution quality to refer to connectivity metric throughout this paper.

\section{Related Work}\label{sec:related}

For a general overview on graph and hypergraph partitioning, there are multiple surveys that provide an introduction to the topic~\cite{PAPA-SURVEY, GRAPH-SURVEY, GRAPH-RECENT-SURVEY, HYPERGRAPH-SURVEY}.
Modern high-quality (hyper)graph partitioners are mostly based on the multilevel paradigm.
Mt-KaHIP~\cite{MT-KAHIP}, Mt-Metis~\cite{MT-METIS} and KaMinPar~\cite{deep-mgp} are some of the most notable publicly available shared-memory graph partitioners, but lack support for hypergraphs.
BiPart~\cite{BIPART} is a deterministic shared-memory hypergraph partitioner.
Mt-KaHyPar~\cite{mt-kahypar-d, mt-kahypar-q, mt-kahypar-journal} is a shared-memory hypergraph partitioner that includes a deterministic mode~\cite{mt-kahypar-sdet} and flow-based refinement~\cite{mt-kahypar-flows}.
Sequential hypergraph partitioners include PaToH~\cite{PATOH} , hMetis~\cite{HMETIS} and KaHyPar~\cite{KAHYPAR-K}.

\paragraph{Parallel Refinement}

For the refinement, partitioning algorithms use a wide range of local search techniques~\cite{MT-KAHIP, MT-METIS, mt-kahypar-d}.
\emph{Label propagation} refinement visits all vertices in parallel and moves them greedily to the block with highest positive gain~\cite{LABEL_PROPAGATION}.
The \emph{Fiduccia-Mattheyses (FM)} algorithm uses priority queues to always apply the move with highest gain, but additionally allows negative gain moves~\cite{FM}, reverting to the best observed solution in the sequence of applied moves.
This allows FM refinement to escape local minima and thereby find higher quality solutions.
However, the FM algorithm is difficult to parallelize because of the serial move order --
some partitioners use the straightforward approach of performing 2-way FM refinement on independent block pairs~\cite{SCOTCH, KAPPA}, but this provides only limited parallelism.
The $k$-way FM variant of Mt-KaHIP and Mt-KaHyPar achieves better parallelism by relaxing the serial order of the moves, using parallel localized searches that are then combined~\cite{MT-KAHIP, mt-kahypar-d}.
Unfortunately, due to its asynchronous and localized nature there is no known way of making this FM variant deterministic while keeping its scalability.

Recently, Gilbert et al.\ introduced the \emph{Jet} refinement algorithm for GPU partitioning~\cite{Jet-Journal}.
Similar to label propagation refinement, Jet is based on rounds of greedy local moves.
However, it allows to include negative gain moves and employs an \emph{afterburner} step to filter the resulting move set in a way that approximates the move sequences computed by FM refinement.
Moreover, Jet allows large temporary imbalances which are repaired in a separate rebalancing step.
This allows a synchronous parallelization well suited for both the GPU and deterministic implementations.
In addition, it gives Jet strong capabilities to escape local minima, consequently achieving high solution quality~\cite{Jet-Journal}.
Maas et al.\ observed that this general approach of \emph{unconstrained refinement}, i.e., the interplay of balance violating moves and rebalancing, can also improve the performance of other local search techniques such as FM refinement~\cite{mt-kahypar-ufm}.\\

A major limitation of local search techniques based on greedy moves is that their decisions consider only local information, i.e., the neighborhood of the moved vertex.
Since global minima can depend on complex interactions between many vertices, this local view might not suffice to find non-trivial improvements.
\emph{Flow-based refinement} was proposed by Sanders and Schulz~\cite{KAFFPA} to enable a more global view on the partitioning problem.
The idea is to compute the minimum cut in a region around the boundary of  two blocks of the partition.
Today, flow-based refinement is an essential component of the highest quality graph and hypergraph partitioning algorithms~\cite{kahypar-hfc, KAHYPAR-MF, mt-kahypar-flows}.
It is considered the strongest local search technique for hypergraph partitioning, but also requires substantially higher running times~\cite{mt-kahypar-journal}.

\paragraph{Deterministic Parallel Partitioning}

To the best of our knowledge, the only existing deterministic parallel partitioners are BiPart~\cite{BIPART} and Mt-KaHyPar-SDet~\cite{mt-kahypar-sdet}, the deterministic mode of Mt-KaHyPar.
BiPart is motivated by applications in VLSI design and uses recursive bipartitioning to obtain partitions with more than two blocks.
Recursive bipartitioning recursively divides the blocks with a two-way partitioning procedure until a $k$-way partition is obtained.
It is simple to implement since refinement algorithms only need to work on two blocks, but it also often produces substantially worse solution quality than direct $k$-way partitioning, which considers all blocks at once~\cite{SimonTeng97}.
In contrast, Mt-KaHyPar-SDet~\cite{mt-kahypar-sdet} implements direct $k$-way partitioning and further techniques to improve quality, including a preprocessing step based on community detection as proposed by Heuer and Schlag~\cite{KAHYPAR-CA}.
As BiPart and Mt-KaHyPar-SDet only use label propagation refinement, they are unable to match the solution quality of FM refinement, let alone flow-based refinement.

\section{Deterministic Jet Refinement}\label{sec:detjet}

The recently proposed Jet refinement algorithm~\cite{Jet-Journal} is designed for the GPU with the goal of matching the quality of FM refinement.
Jet is based on synchronous rounds of local moves, but also introduces a filtering step called the \emph{afterburner} which can be understood as a synchronous approximation of the move priorities used by the FM algorithm.
Combined with allowing temporary weight imbalances, this gives Jet strong capabilities for escaping local minima.
Its synchronous nature makes Jet amenable to a deterministic implementation while enabling superior solution quality to the label propagation techniques used in previous work.

In the following, we present an efficient generalization of the Jet algorithm to hypergraphs as well as techniques to both make the rebalancing deterministic and improve the resulting quality.

\review{The paper mainly focuses on how refinement can be improved to enable determinism, high-quality, and a reasonable performance at the same time. It would be nice if the paper can present the partitioning algorithm at a high-level to help understand how the new algorithms fit in the big picture. For example, for Algorithm 1, while I roughly understand the intuition, I hope I could know more precisely how this step is used in the partitioning algorithm, and how the parameters in the interface (e.g., the moving candidate M, and the target blocks T) are generated in the original algorithm.}

\subsection{Overview}\label{sec:jet-overview}

\begin{algorithm2e}[bt]
	\caption{Jet Overview}
	\label{alg:jet-overview}
	
	\DontPrintSemicolon
	\SetFuncSty{textsc}
	\SetKwFor{ForAll}{forall}{do}
	
	\SetKwFor{DoParallel}{for}{do in parallel}
	
	\While(){round $<$ limit } { 
		  $M \gets $ \FuncSty{ComputeMoveCandidates()}\;
		  $M \gets$ \FuncSty{Afterburner($M$)}\;
	  	\FuncSty{ApplyMoves($M$)}\;
	  	\If(){partition imbalanced}{
	  		\FuncSty{ApplyMoves(RebalancingMoves())}\;
	  	}
		round $\pluseq 1$ \;
		\If(){partition quality improved}{
		  round $\gets 0$ \;
		}

	}
\end{algorithm2e}

Algorithm~\ref{alg:jet-overview} shows a high-level overview of Jet.
One Jet iteration consists of a round of unconstrained moves (i.e., moves that might violate the balance constraint) followed by a rebalancing step~\cite{Jet-Journal}
-- a scheme that has been observed to be effective for overcoming local minima~\cite{mt-kahypar-ufm}.
Initially, Jet selects a set $M \subseteq V$ of move candidates with associated target blocks as follows.
We determine for each vertex $v \in V$ the target block $t(v)$ with highest gain, ignoring partition balance for now.
The gain computation is with respect to the current state of the partition, assuming none of the other vertices move.
This synchronicity is the property that makes Jet amenable to an efficient deterministic implementation.
$M$ may include moves that worsen the objective function, but we use a threshold to exclude moves where the gain is overly negative in relation to the vertex's affinity to its current block.
Given $v \in V_s$, we add $v$ to $M$ only if $\mathrm{gain}(v, t(v)) \ge - \tau \sum_{e \in I(v) : |e \cap V_s| > 1} \omega(e)$ 
for a temperature parameter $\tau \in [0, 1]$.\footnote{If a different objective function than connectivity is used, the inequality needs to be adapted accordingly.}


Afterwards, the afterburner filters $M$ so that only the most promising moves are executed.
We assume a move execution order of $M$, by highest gain first -- analogous to the move order of the FM algorithm. Note that we do not materialize the sorted order.
For each $v \in M$, the afterburner recomputes the gain of moving $v$ to its previously designated target block $t(v)$, assuming all vertices prior to $v$ in the order have been moved to their designated target. Moves with non-positive recomputed gain are filtered out.


If the partition is imbalanced after executing the moves selected by the afterburner, we apply a rebalancing algorithm.
Note that this step is of critical importance for the overall performance of Jet.
Since the unconstrained moves can cause large imbalances in practice, we not only need to reliably restore the balance but also keep the penalty on the objective function incurred by the rebalancing as low as possible.
For this, we replace the original Jet rebalancing algorithm with a newly designed deterministic algorithm.

Moreover, Jet employs \emph{vertex locking} and \emph{rollbacks} to avoid oscillations and decreasing solution quality.
We use the same techniques but refer to the original Jet publication for technical details~\cite{Jet-Journal}.

\subsection{An Efficient Afterburner for Hypergraphs}

\begin{algorithm2e}[bt]
	\caption{Hypergraph-Afterburner}
	\label{alg:hyp-ab-improved}
	
	\DontPrintSemicolon
	\SetFuncSty{textsc}
	\SetKwFor{ForAll}{forall}{do}
	
	\SetKwFor{DoParallel}{for}{do in parallel}
	
	\KwIn{Move candidates $M$, target blocks $t : V \to [k]$}
	recomputed-gain$[v] \gets 0 \quad \forall v \in V$ \;
	\DoParallel{$e \in E$}{
		$\varphi_e[i] \gets |e \cap V_i|  \quad \forall i \in [k]$ \tcp{initial pin counts}
		$L \gets e \cap M$ \;
		sort $L$ by gain$(v, t(v))$  \tcp{descending} 
		\For{pin $v \in L$ in sorted order} {
			$\varphi_e[\Partition(v)] \minuseq 1$\;
			\If(){$\varphi_e[\Partition(v)] = 0$} {
				recomputed-gain$[v] \underset{\text{atomic}}{\pluseq} \omega(e)$ \;
			}
			$\varphi_e[t(v)] \pluseq 1$\;
			\If(){$\varphi_e[t(v)] = 1$} {
				recomputed-gain$[v] \underset{\text{atomic}}{\minuseq} \omega(e)$ \;
			}
		}
	}
	
	\Return $\{ v \in M \mid \text{recomputed-gain}[v] > 0\}$ \;
\end{algorithm2e}


For hypergraphs, it is crucial to avoid algorithms that scale poorly with large hyperedges.
Specifically, computing gains naively by iterating over the pins of incident hyperedges for all vertices is quadratic in the hyperedge size (for each pin $v \in e$, all of $e$ is considered).
When computing the candidate set $M$, this is efficiently solved
by tracking for each hyperedge $e$ the set of incident blocks $\conset(e)$
and the pin count $\varphi_e[i] = |e \cap V_i|$ for each block $V_i \in \conset(e)$~\cite{mt-kahypar-d, mt-kahypar-journal}.
For the afterburner, however, considering the move execution order of $M$ (implicitly sorted by precomputed gain) is crucial to achieve a quality improvement.
For this, we iterate over the hyperedges in parallel and first sort the pins of each hyperedge $e$ according to the order of $M$.
We then simulate the moves for hyperedge $e$ by iterating over the sorted pins
while updating $\varphi_e$ to reflect the partition at the current state, i.e., after moving the current pin $v$ to $t(v)$.
Thereby, for each moved pin $v \in e \cap M$, we can use $\varphi_e$ to calculate the hyperedge's contribution to the gain of the according move.
We accumulate these gain contributions for each vertex $v$,
using atomic fetch-add instructions to avoid race conditions from contributions of different hyperedges.
Algorithm~\ref{alg:hyp-ab-improved} shows the procedure in pseudocode.


While this reduces the afterburner running time per hyperedge $e$ to $\Oh{|e| \log |e|}$, sorting all hyperedges is still expensive in practice.
We further improve the efficiency by only considering the pins in $e \cap M$ and skipping hyperedges where this set is empty.
Additionally, we use specialized implementations for $|e \cap M| \in \{1, 2, 3\}$, which are common cases in practice.


\subsection{Deterministic Rebalancing}

Our rebalancing is inspired by the weak rebalancing of Jet~\cite{Jet-Journal}, but differs in the prioritization and selection of the moves.
The algorithm works in rounds, with each round moving vertices from overloaded blocks to their preferred target block.
To select which vertices are moved, we determine priorities within each overloaded block as follows.
For each vertex $v$, we only consider blocks as target which are still balanced when adding $v$.
Let $\mathrm{gain}(v)$ be the maximum gain for a valid target block.
We define the priority of $v$ as $\mathrm{gain}(v) / c(v)$ if $\mathrm{gain}(v) < 0$ and $\mathrm{gain}(v) \cdot c(v)$ if $\mathrm{gain}(v) \ge 0$.
As the gain is usually negative, this avoids worsening the partition quality more than necessary.
Note that unlike Jet rebalancing, our prioritization includes the vertex weight -- which significantly reduces the penalty incurred by rebalancing~\cite{mt-kahypar-ufm}.

To select the executed moves, Jet uses a bucket-based partial ordering.
This does not work for a deterministic algorithm, since it introduces non-determinism when selecting a subset of vertices from the final bucket.
Instead, we use a parallel sort followed by a parallel prefix sum and binary search to select a minimal set of vertices restoring the balance of the block.
All vertices are synchronously moved to their preferred target block, possibly causing the target to become overloaded.

In order to avoid oscillations caused by newly overloaded blocks, we use the following techniques from the Jet rebalancing algorithm~\cite{Jet-Journal}.
We define a \emph{deadzone} below the maximum allowed block weight $L_{\max}$, with size $d \cdot \varepsilon \lceil c(V) / k \rceil$ for a tuning parameter $d$ (chosen as $d = 0.1$ based on preliminary experiments \nikolai{TODO}).
A block $V_i$ which is not overloaded but within the deadzone, i.e., $c(V_i) \ge L_{\max} - d \cdot \varepsilon \lceil c(V) / k \rceil$, is also not eligible as a target block.
After vertices are moved from an overloaded block, the block is usually within the deadzone and thus can not become overloaded again,
ensuring the overall progress of the algorithm.
Moreover, vertices $v$ with weight larger than $\frac{3}{2} (c(\Pi(v)) - \lceil c(V) / k \rceil)$ are never moved by the rebalancer as they decrease the source block weight below average block weight, increasing the likelihood of oscillations~\cite{Jet-Journal}.



\section{Deterministic Flow-Based Refinement}

Flow-based refinement is currently considered the most powerful iterative improvement technique in terms of solution quality (at the cost of high running time), as it makes optimization decisions based on a more global view~\cite{KAFFPA, KAHYPAR-MF, REBAHFC, kahypar-hfc, mt-kahypar-flows}.
Fortunately, it lends itself well to a deterministic implementation.
Flow-based refinement works on bipartitions, so that $k$-way partitions are refined by scheduling pair-wise refinements on the blocks.
Thus, there are two sources of parallelism that require determinism: scheduling multiple block pairs to run in parallel and parallelizing the max flow computations in each block-pair refinement~\cite{mt-kahypar-flows}.

\review{The description of the flow-based algorithms is very high-level. Some pseudocode/diagrams may help understand the idea better.}

\subsection{Two-Way Refinement}\label{sec:flows:2way}

An easy way to achieve determinism in two-way refinement would be to make the flow algorithm deterministic.
However, it is unclear how to achieve this without sacrificing performance~\cite{baumstark-parallel-flow}.
Perhaps surprisingly, we show that we can work with a non-deterministic flow algorithm. 
Assuming non-deterministic flow assignments, the parallel two-way refinement routine of Ref.~\cite{mt-kahypar-flows} exhibits two sources of non-determinism.
They occur in what is called the \emph{piercing step}.
Below, we give a high-level description of the algorithm to introduce the notion of piercing.
We discuss low-level details for the non-determinism sources and otherwise refer to the original paper~\cite{mt-kahypar-flows} for details.

\paragraph{Algorithm Overview}

To refine a bipartition $(V_1, V_2)$, we work on the induced subhypergraph $H[V_1 \cup V_2]$. We determine initial terminal sets $S \subset V_1$ (the source) and $T \subset V_2$ (the sink). To compute $S$, we perform a BFS restricted to $V_1$ from the boundary vertices until the weight of visited vertices exceeds a threshold (we omit the details here and refer to~\cite{KAFFPA, KAHYPAR-MF, kahypar-hfc, mt-kahypar-flows}). The visited vertices are eligible to change their block assignment from $V_1$ to $V_2$, whereas the not visited vertices constitute $S$. Similarly, we construct $T$ by restricting the search to $V_2$.

\begin{algorithm2e}
	\caption{Incremental Bipartitioning}\label{pseudocode:flowcutter}
	\While() {true} {
		augment flow to maximality regarding $S,T$ \;
		derive source- and sink-side cut $S_r, T_r \subset V$\;
		\If() {$(S_r, V \setminus S_r)$ or $(V \setminus T_r, T_r)$ balanced} {
			\Return balanced partition
		}
		\If() {$c(S_r) \leq c(T_r)$} {
			$S \gets S_r \cup \FuncSty{selectPiercingNode()}$
		} \Else() {
			$T \gets T_r \cup \FuncSty{selectPiercingNode()}$
		}
	}
\end{algorithm2e}

\begin{figure*}[t]
    \hspace{.06\linewidth}
    \begin{minipage}{.37\linewidth}
        \includegraphics[width=\linewidth, page=1]{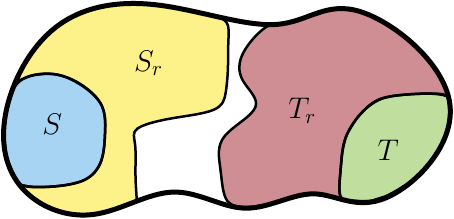}
    \end{minipage}
    \hfill
    \begin{minipage}{.37\linewidth}
        \includegraphics[width=\linewidth, page=2]{illustrations/flowcutter.pdf}
    \end{minipage}
    \hspace{.06\linewidth}
    \caption{
        Venn diagram style visualization of the vertex sets during flow-based refinement. Flow augmentation and computing $S_r, T_r$ on the left. Adding $S_r$ to $S$ and piercing the source-side cut on the right. $S$ in blue, $S_r \setminus S$ in yellow, $T$ in green, $T_r\setminus T$ in red, $V \setminus (S_r \cup T_r)$ in white.
        Taken from Ref.~\cite{REBAHFC} with minor adaptations.
    }\label{fig:illuHFC}
\end{figure*}

As shown in Algorithm~\ref{pseudocode:flowcutter} we then solve a sequence of incremental maximum flow problems whose corresponding minimum $S$-$T$-cuts induce increasingly balanced bipartitions with increasing cut-size~\cite{flowcutter}.
First, we augment the prior flow (initially zero) to be maximal with respect to the current $S$ and $T$. Subsequently, we extract the $S$-reachable vertices $S_r$ and $T$-reachable vertices $T_r$ via (reverse) BFS in the residual hypergraph.
These induce two bipartitions $(S_r, \overline{S_r})$ and $(\overline{T_r}, T_r)$. If either is balanced, we terminate the refinement and replace the old bipartition $(V_1, V_2)$ if the new cut is smaller (or equal and has smaller imbalance).
Otherwise, we continue refinement. We transform all vertices on the smaller side to a terminal, i.e., $S \gets S_r$ if $c(S_r) \leq c(T_r)$ and $T \gets T_r$ otherwise.
To find a different cut in the next iteration, we add one more vertex (the piercing vertex) to the smaller side.
Thus, the bipartitions contributed by the smaller side will be more balanced in future iterations.
We provide a Venn diagram style illustration of the vertex sets in Figure~\ref{fig:illuHFC}.

\paragraph{The Bipartitions are Deterministic}

The employed parallel maximum flow algorithm~\cite{baumstark-parallel-flow} is non-deterministic.
Interestingly however, there are two unique inclusion-minimal (respectively maximal) min-cut bipartitions for a fixed $S$ and $T$; even under different maximum preflow assignments.
These are precisely the two min-cut bipartitions we inspect. As they are unique, the refinement is deterministic, even though the flow algorithm is not.
All minimum $S$-$T$ cuts are nested in a DAG by vertex inclusion dependencies~\cite{PicardQ82}.
Let $(X,\overline{X})$ and $(Y, \overline{Y})$ be minimum $S$-$T$ cuts.
Then $(X \cup Y, \overline{X \cup Y})$ and $(X \cap Y, \overline{X \cap Y})$ are also minimum $S$-$T$ cuts.
The two bipartitions we inspect are the inclusion-minimal source side (intersection of all min-cut source-sides/union of sink-sides) and inclusion-maximal source side (union of all min-cut source-sides/intersection of sink-sides). By virtue of minimality (resp. maximality) they are unique and thus deterministic.
For proof details, we refer to the paper of Picard and Queyranne~\cite{PicardQ82}.


\paragraph{Piercing Vertex Selection}

We select a piercing vertex from the \emph{boundary vertices} of the new min-cut, which is the set of vertices incident to cut hyperedges.
While extracting the new min-cuts via BFSs in the residual hypergraph, we discover new boundary vertices as piercing candidates, which we insert into an array in the order they are discovered.
While the set of visited vertices is deterministic, the visit order is non-deterministic, as different hyperedges may be residual with different flow assignments.
To ensure a deterministic order, we sort the newly inserted candidates a posteriori by vertex ID, and then select a piercing vertex from the array.
We refer the interested reader to Ref.\cite{kahypar-hfc} for details on the selection criteria.


\paragraph{Termination Checks}

The second source of non-determinism stems from the interplay of two implementation details. 
1) When piercing a vertex $u$ with excess flow $\mathrm{exc}(u) > 0$ into the sink terminal set $T \gets T \cup \{u\}$, its excess must be added to the flow value $f \gets f + \mathrm{exc}(u)$, as that flow is routed from $S$ to $u$, which is now a sink.
The flow value can increase outside a flow computation!
However, the excess values are non-deterministic under non-deterministic preflows.
2) We terminate the refinement if the flow value becomes equal to the cut of the old bipartition, as we are not interested in larger cuts.
With equal cut value we may accept the new bipartition since it may have smaller imbalance.

The non-determinism occurs when there is a cut with equal value, yet in one run the flow value is reached through piercing and in another it is reached through flow computation.
So far, the termination check happens after piercing, before augmenting the flow.
If it triggers after piercing and thus flow computation is skipped, we do not output a possible cut with equal flow value, as we need to run flow computation to check whether the flow is maximal.
If the value is reached through flow computation, we output the cut.
The two scenarios lead to different outcomes and thus non-deterministic behavior.

We fix this by performing the termination check before piercing.
In both of the scenarios above we thus run flow computation. In the first, it detects that the flow is already maximal from the augmentation during piercing, in the second the augmentation happens during flow computation.
The impact of this change on running time is small, as there is also a termination check in the flow computation which will trigger as soon as the bound is exceeded.

We note that this is a very subtle and hard-to-detect bug.
It also leads to missed solutions in the non-deterministic setting, though we note that the bug is harmless in that setting.

\subsection{K-Way Refinement: Matching Schedule}\label{sec:flow-scheduling}

To refine a $k$-way partition, we schedule multiple two-way refinements in parallel, following the \emph{active block scheduling} strategy of Sanders and Schulz~\cite{KAFFPA, mt-kahypar-flows}.
The high-level idea of this round-based schedule is to skip block pairs where neither block contributed to an improvement in the previous round, marking them as inactive. Block pairs where both blocks are inactive are not scheduled.
In Mt-KaHyPar, the same block can participate in multiple concurrent two-way refinements~\cite{mt-kahypar-flows}. Conflicts from moving the same vertex more than once are resolved by skipping vertices which are not assigned to the expected block, i.e., moves are applied on a first-come-first-serve basis, which is non-deterministic.

Therefore, we use a schedule where each block only participates in one refinement at a time, which comes at the expense of less parallelism in the scheduler.
The quotient graph $Q=(V_Q, E_Q)$ has the blocks as vertices $V_Q = \{V_1, \dots, V_k\}$ and edges between two blocks that are connected through a cut hyperedge; $E_Q = \{ (V_i, V_j) \mid \exists e \in E: e \cap V_i \ne \emptyset \text{ and } e \cap V_j \ne \emptyset \}$.
We schedule a maximal matching in $Q$ in parallel, and synchronize before scheduling the next one, until all quotient graph edges have been scheduled.
To combat stragglers (small matchings in late iterations), we sort the blocks by their degree in the remaining quotient graph of unscheduled edges, preferring to schedule edges involving high degree blocks first.

\section{Improving Deterministic Coarsening}

In the coarsening phase, we repeat two alternating steps until the current hypergraph is \emph{sufficiently small for initial partitioning}\cite{kahypar-journal}.
First we cluster vertices using a local moving procedure with the heavy-edge-rating function~\cite{PATOH}.
Then we contract each cluster to form a new vertex. 
We build upon the existing synchronous coarsening algorithm of Ref.~\cite{mt-kahypar-sdet}, reusing the contraction algorithm as is.
We propose two small improvements: a prefix-doubling schedule of synchronous subrounds and preventing accidental vertex swaps.
Additionally, we fix a bug in the rating function computation of the original implementation (see below), leading to a noticeable improvement in downstream solution quality.
We refer to Appendix~\ref{app:coarsening-ablations} for an ablation study, which evaluates the impact of each change.

At the start of the clustering algorithm, each vertex is in a singleton cluster by itself.
We parallelize over the vertices.
For each vertex $u$ still left in a singleton cluster, we compute the \emph{heavy-edge-rating function} $\sum_{e \in \incnets(u)} \frac{\omega(e) \cdot 1_{|e \cap C| > 0}}{|e|-1}$ over clusters $C$ in its neighborhood and select the highest rated cluster~\cite{PATOH}, subject to a loose weight constraint.
Heavy edge rating prefers hyperedges with large weight, as they have a stronger effect on the partitioning objective.
Additionally small hyperedges are preferred, as they are more easily removed through contraction (when reaching $|e| = 1$).
The vertex weight constraint ensures that the coarse hypergraphs have a balanced partition which can be found by greedy algorithms.

The previous implementation of the rating erroneously added the weight of an hyperedge multiple times (per pin in the target cluster) instead of only once.
Fixing this bug improves the clustering decisions.

\begin{algorithm2e}[t]
	\SetEndCharOfAlgoLine{}
	\caption{Synchronous Coarsening}\label{algo:sync-local-moving}
	Split $V$ into disjoint sets $B_1, ..., B_r$ \;
	$T[u] \gets \bot \quad \forall u \in V$ \tcp{target clusters}
	\For() {$i=0$ \KwTo $r$} {
		\ParallelFor() {$u \in B_i$} {
			$T[u] \gets \underset{j \neq u}{\argmax} \sum_{e \in \incnets(u)} \frac{\omega(e) \cdot 1_{|e \cap C_j| > 0}}{|e|-1}$ 
		}
		$\mathcal{C} \gets \{T[u] \mid u \in B_i, T[u] \neq \bot\}$ \;
		$M[j] \gets \{ u \in B_i \mid T[u] = j \} \;  \forall j \in \mathcal{C} $ \tcp{group-by}
		\ParallelFor(){$j \in \mathcal{C}$}{
			sort $M[j]$ by increasing weight and ID \;
			$W[l] \gets \sum_{u \in M[j][0:l]}c(u)$ \tcp{prefix sum}
			pos $\gets$ binary search $($max cluster weight $-$ current weight of cluster $j)$ in $W$ \;
			move $M[j][0:\text{pos}]$ to cluster $j$ \;
		}
	
	}
\end{algorithm2e}

In non-deterministic mode, a vertex joins its preferred target cluster immediately, i.e., vertices are moved asynchronously.
In deterministic mode, we follow the structure of Algorithm~\ref{algo:sync-local-moving}. 
We first compute the target clusters for a subset $B_i \subset V$ without moving $B_i$.
Since the new cluster weights are not known during target cluster selection,
we then need a separate step to determine a a subset of $B_i$ which can be moved without violating the cluster weight constraint.
For this, we first group moves by their target cluster.
For each target cluster $j$, we then move as many vertices as possible within the cluster weight constraint, with preference to lower weight vertices.
We synchronize after each subround, so that move decisions in future subrounds can be informed by changed cluster labels.
We perform multiple sub-rounds untill all vertices were considered.

\paragraph{Prefix Doubling Schedule}

In the existing deterministic coarsening~\cite{mt-kahypar-sdet}, the vertices in a synchronous sub-round $B_i$ are determined by randomly splitting into sets of roughly equal size and the number of sub-rounds is small ($r=3$ proved best in terms of solution quality).
During the first sub-round of the synchronous algorithm, all clusters still appear as singletons.
This is contrary to asynchronous coarsening, where clusters start to form quickly and these changes inform the move decisions of other vertices.

To incorporate early aggregations into the move decisions of the synchronous algorithm, while keeping synchronization costs reasonable, we employ a prefix-doubling approach to sub-round sizes.
We start with one vertex per sub-round and perform 100 initial sub-rounds.
Notably, these are sequential steps since we employ parallelism over vertices.
In each subsequent sub-round we double the size, and thus the amount of parallelism, until a limit of 1\% of the vertices.

\paragraph{Preventing Vertex Swaps}

A second undesirable side effect in synchronous clustering is when two vertices $u$ and $v$ are part of the same sub-round where $u$ joins $v$'s cluster and vice versa.
Unfortunately, both are left in a cluster without their desired contraction partner. 
Moreover, additional vertices that wanted to join $u$ are now in a cluster with $v$ but not $u$.
To resolve this, we perform an additional pass before the group-by-target-cluster step to detect all such pairs (i.e., pairs with $T[u] = v$ and $T[v] = u$), and merge them by either setting $T[u] \gets u$ or $T[v] \gets v$.
If other vertices joined $u$ or $v$, we pick the cluster with larger weight as the target.
Note that the propositions $T$ are still subject to approval for the cluster weight constraint.


\section{Experimental Evaluation}

In the following, we evaluate the parameters and overall performance of our algorithms.
In Section~\ref{sec:exp-coarsening}, we evaluate the proposed improvements to deterministic coarsening relative to the algorithm currently used in Mt-KaHyPar~\cite{mt-kahypar-sdet}.
We provide a detailed discussion of the parameter choices that enable Jet refinement to perform well on both graphs and hypergraphs in Section~\ref{sec:exp-jet-parameters} and present strong scalability results for Jet in Section~\ref{sec:exp-scalability}.
Finally, in Section~\ref{sec:exp-comparison}, we compare the resulting high-quality deterministic algorithm with the state of the art in both deterministic and non-deterministic partitioning.
This includes the non-deterministic parallel multilevel graph partitioners Mt-Metis~\cite{MT-METIS} (with hill-scanning and two-hop coarsening) and KaMinPar~\cite{deep-mgp} (default configuration),
the parallel multilevel hypergraph partitioner Mt-KaHyPar in both deterministic mode~\cite{mt-kahypar-sdet} and non-deterministic mode~\cite{mt-kahypar-d, mt-kahypar-ufm, mt-kahypar-flows} (default configuration with unconstrained FM, as well as quality configuration with flow-based refinement),
and the deterministic parallel hypergraph partitioner BiPart~\cite{BIPART}.
We exclude sequential partitioners hMetis, PaToH and KaHyPar from the comparison, since they are outperformed by Mt-KaHyPar.

\subsection{Setup}\label{sec:exp-setup}

Our algorithms are implemented within the Mt-KaHyPar framework, reusing the existing implementations for deterministic preprocessing and initial partitioning while adapting the coarsening and replacing the refinement algorithms.
The source code is available at \url{https://github.com/kahypar/mt-kahypar/tree/rk-nm/new-deterministic}.

The majority of the experiments are run on an AMD EPYC 9684X (one socket with 96 cores), clocked at 2.55-3.7\,GHz with 1536\,GB RAM and 1152\,MB L3 cache.
Due to time constraints the scaling experiments were performed on different hardware.\footnote{
    AMD EPYC Rome 7702P, 64 cores, frequency 2.0-3.35\,GHz with 1024GB RAM and 256MB L3 cache.
}
We run all algorithms with 64 cores.

\paragraph{Bechmark Sets}
We use three different benchmark sets from the literature to represent a diverse set of possible inputs.
First, we use a set of 94 hypergaphs compiled by Gottesbüren et al.~\cite{mt-kahypar-d}\footnote{
    Publicly available at \url{https://zenodo.org/records/15386567}
}, with sizes of up to 139 million vertices, up to 138 million hyperedges and up to 1.94 billion pins.
Further, we use the two graph benchmark sets of Maas et al.~\cite{mt-kahypar-ufm}\footnote{
    Publicly available at \url{https://zenodo.org/records/15386627}
}, which are categorized into 38 irregular instances (such as social networks) and 33 regular instances (such as mesh graphs) -- a classification which is useful since the authors observed large differences in the behavior of partitioning algorithms on these two benchmark sets.
The sizes range from 5.4 million edges to 1.8 billion edges for the irregular graphs and 12.7 million edges to 575 million edges for the regular graphs.
More details on the composition of the benchmark sets are presented in Appendix~\ref{app:benchmark-sets}.

\begin{figure}
    \begin{minipage}{\textwidth}
    \ifpdfplots
    \includegraphics{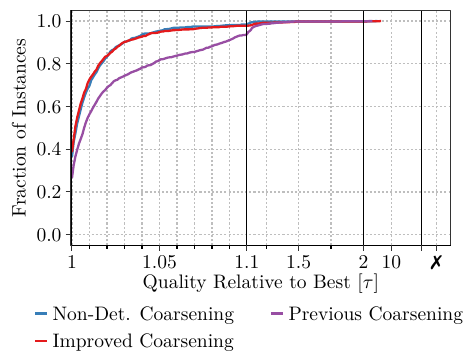}
    \else
    \tikzsetnextfilename{coarsening}%
    \input{tikz/coarsening}%
    \fi

    \end{minipage}
    \vspace{-10pt}
    \caption{Impact of improved coarsening on solution quality, on all instances.}
    \label{fig:coarsening}
\end{figure}

\paragraph{Methodology}
We use imbalance value $\varepsilon = 0.03$, which is a common choice in the literature~\cite{HYPERGRAPH-SURVEY}.
We further use $k \in \{2, 8, 11, 16, 27, 64, 128\}$ and five different seeds for randomization per instance (graph and $k$) when comparing deterministic Jet refinement to the current state of the art, but only three seeds for the remaining experiments due to time constraints.
We aggregate the running time and the objective (the connectivity metric for hypergraphs and edge cut for graphs) for each instance using the arithmetic mean over the seeds.
When aggregating over multiple instances, we use the geometric mean to ensure that each instance has the same relative influence on the result.
For aggregated running times, we include imbalanced partitions and, if a run exceeded the time limit, use the time limit itself in
the aggregate.

\begin{figure*}
    \begin{minipage}{\textwidth}
    \ifpdfplots
    \includegraphics{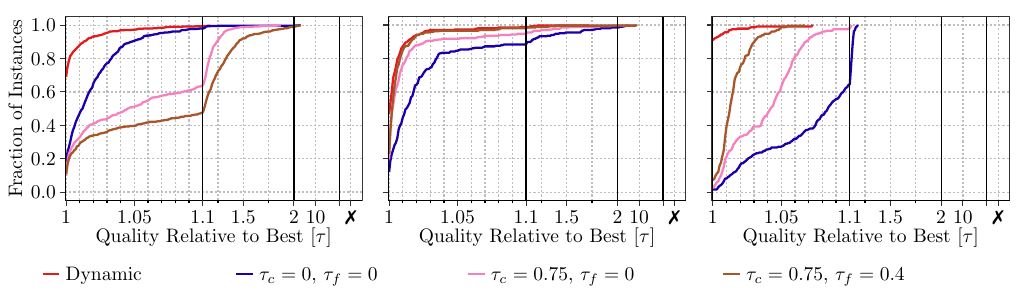}
    \else
    \tikzsetnextfilename{temperatures}%
    \input{tikz/temperatures}%
    \fi

    \end{minipage}
    \vspace{-20pt}
    \caption{Comparison of solution quality for different temperature settings for Jet refinement on hypergraphs (left), irregular graphs (center) and regular graphs (right).
        $\tau_c$ denotes the temperature value used for coarse levels, while $\tau_f$ denotes the value used for the finest level of the input.
        The dynamic configuration uses three rounds with values between $0.75$ and $0$.}
    \label{fig:temperatures}
\end{figure*}

\begin{figure}
    \begin{minipage}{\textwidth}
    \ifpdfplots
    \includegraphics{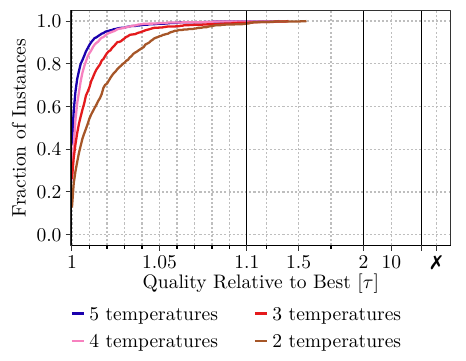}
    \else
    \tikzsetnextfilename{num_temperatures}%
    \input{tikz/num_temperatures}%
    \fi

    \end{minipage}
    \vspace{-10pt}
    \caption{Solution quality on all instances for different numbers of Jet refinement rounds with dynamically decreasing temperature value $\tau$.}
    \label{fig:num_temperatures}
\end{figure}

\begin{figure}
    \begin{minipage}{\textwidth}
    \ifpdfplots
    \includegraphics{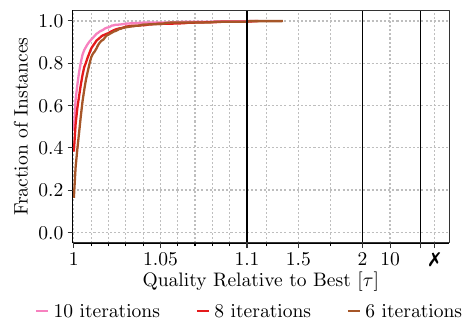}
    \else
    \tikzsetnextfilename{iterations}%
    \input{tikz/iterations}%
    \fi

    \end{minipage}
    \vspace{-10pt}
    \caption{Solution quality on all instances for different values of maximum allowed Jet iterations without quality improvement.}
    \label{fig:iterations}
\end{figure}

\paragraph{Performance Profile Plots}
We compare the quality of multiple different algorithms using \emph{performance profiles}~\cite{PERFORMANCE-PROFILES}.
Let $\mathcal{A}$ be the set of algorithms to compare, let $\mathcal{I}$ be the set of instances and let $q_A(I)$ be the (minimization) objective 
of algorithm $A \in \mathcal{A}$ on instance $I \in \mathcal{I}$.
For each algorithm $A$, we plot on the $y$-axis the fraction of instances where $q_A(I) \le \tau \cdot \min_{A' \in \mathcal{A}} q_{A'}(I)$, with $\tau$ being on the $x$-axis.
An algorithm $A$ is considered to perform better when achieving higher fractions at lower $\tau$ values.
For $\tau=1$, the $y$-value indicates the fraction of instance where $A$ found the best solution.
If an algorithm timed out or could not produce a balanced partition for \emph{every} run on a given instance, we mark the instance with \ding{55}.

\subsection{Deterministic Coarsening}\label{sec:exp-coarsening}

In Figure~\ref{fig:coarsening}, we evaluate our improved coarsening by comparing with the existing (non-)deterministic coarsening.
All versions use DetJet refinement.
We can observe that there is a notable quality gap from the baseline deterministic coarsening to the non-deterministic coarsening used by Mt-KaHyPar-Default (1.5\% in the geometric mean).
The difference is most pronounced on hypergraphs; on graph instances it is only 0.6\%.
Our improved coarsening fully closes the quality gap, resulting in identical solution quality to the non-deterministic coarsening.
We evaluate the specific contribution of each change in detail in Appendix~\ref{app:coarsening-ablations}.
The largest improvement is due to the bugfix, followed by preventing vertex swaps.
Prefix doubling does not improve solution quality of the final partition, but substantially improves quality of the initial partitions.
This suggests that, while it leads to better coarsening decisions, the refinement is strong enough to compensate for bad coarsening decisions.
We keep prefix doubling enabled, as it offers a more robust configuration with equivalent empirical quality.
The previous and improved deterministic coarsening are 16\%, respectively 24\% slower than non-deterministic coarsening in the geometric mean.
This is due to overheads incurred by the approval step and more frequently synchronizing threads.
Note that for most instances, the running time is dominated by the refinement step (see Appendix~\ref{app:running-time-share}). Therefore, an increase in coarsening time has rather small impact on the overall running time.

\subsection{Tuning DetJet Parameters}\label{sec:exp-jet-parameters}

Jet has several parameters with impact on solution quality and running time.
The temperature parameter $\tau$ (see Section~\ref{sec:jet-overview}) determines how many moves with initially negative gain are considered in the afterburner.
As already observed by Gilbert et al., too small values of $\tau$ make it hard to escape local minima, while large $\tau$ may allow moves that actually worsen the objective~\cite{Jet-Journal}.
The authors use a larger value ($\tau_c = 0.75$) on the coarse levels and a small value ($\tau_f = 0.25$) on the top level.
As we show in Figure~\ref{fig:temperatures}, the issue is even more severe for hypergraphs, and their behavior is different from graphs.
$\tau = 0$ consistently gives the best results for hypergraphs while using values up to $0.75$ is crucial on graphs.
Our data indicates that this is due to the afterburner performing worse for hyperedges than for graph edges.
This behavior can be explained as follows.
Any pin of a given hyperedge $e$ might increase the connectivity of $e$ when moved to a different block.
Therefore, approximating the resulting gain is harder for hyperedges, where many interactions are possible, than for graph edges, which connect only two vertices.


To find a configuration that works well across different input types, we follow the suggestion of Sanders and Seemaier~\cite{distributed-jet} to use different temperatures on the same level, decreasing from $\tau=0.75$ to $\tau=0$ in equidistant steps.
As Figure~\ref{fig:temperatures} demonstrates, this approach consistently outperforms the best configuration with a fixed value, though at the expense of increased running time.
Figure~\ref{fig:num_temperatures} shows the effect of the number of different temperatures on solution quality.
We settle on three temperatures to strike a good trade-off between quality and time.
Up to four temperatures, each step achieves a notable improvement in quality.
Yet, diminishing returns set in quickly (0.7\% geometric mean improvement going from 2 to 3, 0.5\% improvement going from 3 to 4), while the geometric mean running time increases significantly (5.23\,s, 5.98\,s, 6.66\,s and 7.31\,s, respectively).
Three temperatures thus provide a good compromise for high quality with a reasonable running time.

\begin{table*}[h]
    \vspace{10pt}
    \centering
    \begin{tabular}{lrrrr}
        Algorithm & Hypergraphs & Irregular Graphs & Regular Graphs & All Instances\\
        \midrule
        \textbf{DetJet} & 5.55 & 10.42 & 3.91 & 5.98\\
        Mt-KaHyPar-Default & 4.72 & 9.21 & 3.52 & 5.19\\
        Mt-KaHyPar-SDet & 2.89 & 4.25 & 1.94 & 2.91\\
        KaMinPar & - & 1.16 & 0.72 & 0.93\\
        Mt-Metis$^*$ & - & 4.10 & 0.99 & 2.12\\
        \textbf{DetFlows}$^{**}$ & 30.25 & 258.26 & 53.41 & 55.54\\
        Mt-KaHyPar-Flows$^{**}$ & 23.03 & 236.98 & 36.47 & 43.19\\
    \end{tabular}

    \vspace{-3pt}
    \caption{Geometric mean running time of the partitioners in seconds, excluding failed runs for Mt-Metis ($^*$) and using the timeout value for instances where flow-based refinement could not complete in time ($^{**}$).}
    \label{tab:running_times}
\end{table*}

A similar trade-off is available through the maximum number of iterations performed without encountering a better partition (see Ref.~\cite{Jet-Journal}).
While the original Jet implementation uses 12 iterations, we can tune this down since using three different temperatures ensures high quality even with less iterations per round.
Figure~\ref{fig:iterations} shows that the impact on quality is small even with only 6 iterations.
We use 8 iterations in our final configuration since this worked best in preliminary experiments, though the evaluation indicates that reducing this even further might actually be preferable.


\subsection{Strong Scaling Experiments}\label{sec:exp-scalability}

\begin{figure}
    \begin{minipage}{\textwidth}
    \ifpdfplots
    \includegraphics{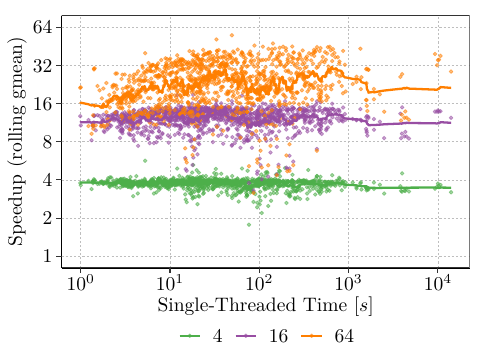}
    \else
    \tikzsetnextfilename{speedup}%
    \input{tikz/speedup}%
    \fi

    \end{minipage}
    \vspace{-12pt}
    \caption{Strong scalability results for our Jet refinement algorithm on all instances.
        Each point corresponds to an instance (sorted by sequential time), and the line is a rolling window geometric mean.
        \nikolai{alte Daten}}
    \label{fig:speedup}
\end{figure}

In Figure~\ref{fig:speedup}, we evaluate self-relative speedups of deterministic Jet.
We compare 4, 16 and 64 cores to a single-threaded run for each instance and show a rolling window geometric mean.
With 4 cores we achieve an almost linear speedup, and still roughly $12\times$ for 16 cores.
Adding even more cores is less effective, with a $22.5\times$ average speedup for 64 cores, though there are also many instances with a speedup of at least $32\times$, with the maximum speedup as $55\times$.



\subsection{Comparison with other Algorithms}\label{sec:exp-comparison}

We compare DetJet (our configuration with Jet refinement) and DetFlows (our configuration with both Jet refinement and flow-based refinement) with the current state of the art, including non-deterministic as well as deterministic partitioning algorithms.
We exclude flow-based refinement from the DetJet evaluation, providing a separate evaluation instead -- due to the substantially higher running times we consider it to be a separate category in the running time versus quality trade-off.

\paragraph{Deterministic Jet Refinement}

\begin{figure*}
    \begin{minipage}{\textwidth}
    \ifpdfplots
    \includegraphics{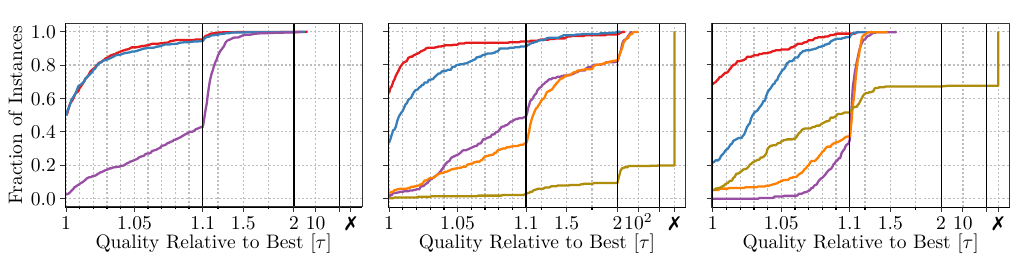}
    \else
    \tikzsetnextfilename{final_comparison}%
    \input{tikz/final_comparison}%
    \fi

         \vspace{-20pt}
    \end{minipage}
    \begin{minipage}{\textwidth}
    \ifpdfplots
    \includegraphics{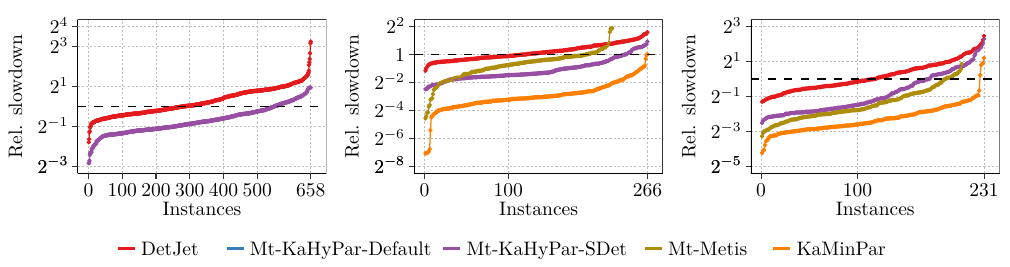}
    \else
    \tikzsetnextfilename{relative_running_time}%
    \input{tikz/relative_running_time}%
    \fi

    \end{minipage}
    \vspace{-10pt}
	\caption{Comparing DetJet to state-of-the-art partitioners on hypergraphs (left), irregular graphs (center) and regular graphs (right). We use performance profiles for the solution quality (top, higher is better) and plot relative running times (bottom, lower is better) in relation to Mt-KaHyPar-Default. Each point corresponds to an instance, which are individually sorted for each algorithm by relative running time. Note that for Mt-Metis, instances where the run failed are excluded from the running time plot.} 
	\label{fig:final-comparison}
\end{figure*}

In Figure~\ref{fig:final-comparison}, we compare the solution quality of DetJet to state-of-the-art parallel partitioners.
DetJet finds the best partition of all considered algorithms on 50\% of the hypergraphs, 61\% of the irregular graphs and 64\% of the regular graphs.
Overall, the quality is equivalent to the non-deterministic mode of Mt-KaHyPar on hypergraphs and slightly better on both types of graphs.
There is a significant gap in quality to all remaining algorithms -- specifically, we achieve an overall improvement over Mt-KaHyPar-SDet of $1.18\times$ in the geometric mean and even of $1.42\times$ if only irregular graphs are considered.
KaMinPar achieves similar quality to Mt-KaHyPar-SDet, which is expected since the default mode of KaMinPar prioritizes speed over quality.
Mt-Metis is often unable to find a balanced solution at all, particularly so on irregular graphs.\footnote{
    On some instances Mt-Metis could not produce a result at all, due to a segfault or running out of main memory -- this is mostly equivalent to previous results on the same instances~\cite{mt-kahypar-ufm}.
}

Figure~\ref{fig:final-comparison} also provides a detailed instance-wise comparison of running times.
The running time of DetJet is similar to Mt-KaHyPar-Default, being 15\% slower in the geometric mean (see Table~\ref{tab:running_times} for average running times)
-- a good trade-off, since DetJet provides both determinism and slightly better quality on graph instances.
Note that on hypergraphs there are a few outliers with a relative slowdown of $4\times$ or more.
The remaining partitioners are faster, with Mt-KaHyPar-SDet being roughly $2\times$ faster than DetJet and Mt-Metis being $2\times$ faster and $4\times$ faster for irregular graphs and regular graphs, respectively.
The fastest algorithm is KaMinPar, with a difference of more than $8\times$ for irregular graphs and $5\times$ for regular
graphs.
However, as discussed, none of these algorithms provide high-quality partitions.

\paragraph{Deterministic Flow-Based Refinement}

\begin{figure}
    \begin{minipage}{\textwidth}
    \ifpdfplots
    \includegraphics{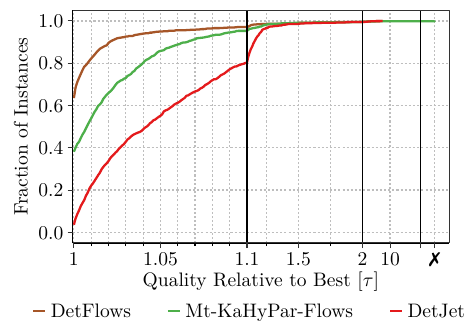}
    \else
    \tikzsetnextfilename{flows_comparison}%
    \input{tikz/flows_comparison}%
    \fi

    \end{minipage}
    \vspace{-10pt}
    \caption{Comparing the solution quality of deterministic flows to non-deterministic flows on all instances, except for three hypergraphs and seven graphs where flow-based refinement could not produce a partition within the time limit of one hour (see Appendix~\ref{app:benchmark-sets}). Additionally, DetJet is included as a baseline.}
    \label{fig:flows_comparison}
\end{figure}

\begin{figure}
    \begin{minipage}{\textwidth}
    \ifpdfplots
    \includegraphics{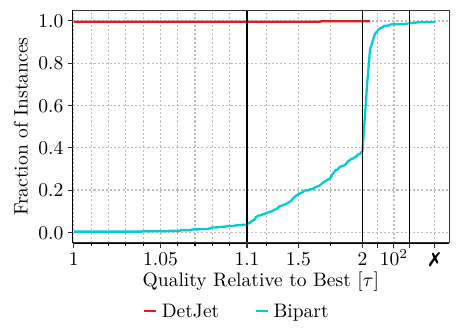}
    \else
    \tikzsetnextfilename{cmp_to_bipart}%
    \input{tikz/cmp_to_bipart}%
    \fi

    \end{minipage}
    \vspace{-10pt}
    \caption{Comparing the solution quality of DetJet to Bipart on hypergraphs.
        \vspace{48pt}}
    \label{fig:cmp_to_bipart}
\end{figure}

While both Jet and FM refinement are capable of providing high quality within a reasonable time,
flow-based refinement computes even better partitions at the cost of a significant increase in running time.
As shown in Figure~\ref{fig:flows_comparison}, both the deterministic and non-deterministic variant provide a notable quality improvement in comparison to DetJet (4-5\% in the geometric mean).
Somewhat surprisingly, DetFlows even outperforms Mt-KaHyPar-Flows by a margin of 1.2\% in the geometric mean.
This small improvement is consistent for hypergraphs and both types of graphs.
We have not identified a root cause yet.
The quality improvement comes at the expense of slower running times, with a $5.45\times$ difference between DetFlows and DetJet on hypergraphs and a $24.8\times$ and $13.7\times$ difference on irregular graphs and regular graphs, respectively (see Table~\ref{tab:running_times}).\footnote{
    The gap between hypergraphs and graphs is because both FM and Jet refinement use a specialized implementation for graphs, with more efficient data structures and algorithms than the hypergraph implementation (compare Ref.~\cite{mt-kahypar-flows}). The flow-based refinement currently uses the same implementation for hypergraphs and graphs, thereby increasing the running time gap on graph instances even further. 
}
In direct comparison to Mt-KaHyPar-Flows, the DetFlows algorithm is 29\% slower in the geometric mean, which is similar on all three instance types.
The majority of this slowdown is explained by the more restricted block pair scheduling strategy, which is required for deterministic results but reduces the available parallelism (see Section~\ref{sec:flow-scheduling}).

\subsection{Individual Comparison with BiPart}

Figure~\ref{fig:cmp_to_bipart} provides a direct comparison on the hypergraph benchmark set to BiPart~\cite{BIPART}, the only previous deterministic parallel partitioner apart from Mt-KaHyPar-SDet.
We were unable to run BiPart ourselves without segmentation faults. Therefore, we perform a separate comparison of DetJet with BiPart using data from Ref.~\cite{mt-kahypar-sdet}, running DetJet on equivalent hardware for a fair comparison.\footnote{
    AMD EPYC Zen 2 7742, 64 cores, frequency 2.25-3.4\,GHz with 1024GB RAM and 256MB L3 cache.
}
Consistent to the previous results~\cite{mt-kahypar-sdet}, we observe that BiPart has vastly inferior solution quality to DetJet, with DetJet finding a better partition for 99.6\% of the instances.
The geometric mean difference in quality is $2.4\times$ and BiPart even produces more than $10\times$ worse partitions on 4\% of the instances.
Furthermore, BiPart is also slower than DetJet by a factor of $3.5$ in the geometric mean.
Therefore, DetJet clearly outperforms BiPart both for solution quality and running time.


\section{Conclusion}

We develop the first deterministic parallel algorithm for hypergraph partitioning that matches the quality of the best non-deterministic and sequential solvers.
Our deterministic Jet refinement achieves a good quality to running time trade-off, with crucial aspects being the efficient handling of large hyperedges, a high-quality rebalancing step and using multiple rounds with a different temperature parameter $\tau$.
Furthermore, we build deterministic flow-based refinement atop a high-performance non-deterministic maximum flow algorithm.
Our experiments demonstrate that our refinement algorithms together with small improvements to deterministic coarsening match or exceed the quality of the state-of-the-art non-deterministic partitioner Mt-KaHyPar with FM refinement and flow-based refinement, respectively.

\newpage


\bibliography{references}

\clearpage

\onecolumn

\appendix

\section{Additional Details on the Benchmark Sets}\label{app:benchmark-sets}

\begin{table}[!h]
	\caption{Graph benchmark sets from Ref.~\cite{mt-kahypar-ufm}, with irregular graphs in the upper half and regular graphs in the lower. The graphs are grouped into classes based on their origin. Within each class, the graphs are ordered by increasing number of edges.}
	\label{tab:graphs}
	
	\vspace{6pt}
	\begin{tabular}{lrp{0.775\textwidth}}
		Class & \# & Graphs\\
		\midrule
		Social & 8 & imdb2021, flickr-und, livejournal, hollywood, orkut, sinaweibo, twitter2010, friendster \\
		Web & 7 & mavi201512020000, indochina2004, arabic2005, uk2005, webbase2001, it2004, sk2005 \\
		Wiki & 5 & eswiki2013, itwiki2013, frwiki2013, dewiki2013, enwiki2022 \\
		Brain & 5 & bn-M87117515, bn-M87123142, bn-M87122310, bn-M87126525, bn-M87128519-1 \\
		Compression & 6 & sources1GB-7, sources1GB-9, english1GB-7, dna1GB-9, proteins1GB-7, proteins1GB-9 \\
		Artificial & 7 & rmat-n16m24, kron-g500n19, rhg-n23d4, kron-g500n20, kron-g500n21, rhg-n23d20, rmat-n25m28 \\
		\midrule
		Road & 2 & asia-osm, europe-osm \\
		Biology & 6 & cage15, kmer-V2a, kmer-U1a, kmer-P1a, kmer-A2a, kmer-V1r \\
		Optimization & 2 & nlpkkt200, nlpkkt240 \\
		Finite element & 16 & ldoor, afshell10, boneS10, Hook1498, Geo1438, Serena, audikw, channelb050, LongCoup-dt6, dielFilterV3, MLGeer, Flan1565, Bump2911, CubeCoup-dt6, HV15R, Queen4147 \\
		Semiconductor & 4 & nv2, vas-stokes2M, vas-stokes4M, stokes \\
		Artificial & 3 & delaunay-n24, rgg-n263d, rgg-n26
	\end{tabular}
\end{table}

\paragraph{Graph Instances}
The graphs in Table~\ref{tab:graphs} were compiled from the Network Repository~\cite{networkrepository}, the SuiteSparse Matrix Collection~\cite{SPM} and the Laboratory for Web Algorithmics (LAW)~\cite{WebInstances1, WebInstances2, WebInstances3}.
Specifically, this includes a set of graphs modeling the human brain~\cite{BrainInstances}, the Flickr social network~\cite{SocFlickrInstance} and the Sina Weibo social network~\cite{SinaWeiboInstance}.
In addition, artificial graphs from a variety of random graph models are included.\footnote{
The graph instances as well as our experimental results are available here: \url{https://zenodo.org/records/15386627}
}

\paragraph{Hypergraph Instances}
The 94 instances in the hypergraph benchmark set compiled by Gottesbüren et al.~\cite{mt-kahypar-d} stem from three sources: the DAC 2012 Routability-Driven Placement Contest~\cite{DAC}, the SuiteSparse Matrix Collection~\cite{SPM}, and the International SAT Competition 2014 \cite{SAT14}.\footnote{
The hypergraph instances are available here: \url{https://zenodo.org/records/15386567}
}

\paragraph{Excluded Instances for Flow-Based Refinement}
The following instances were excluded from the evaluation of flow-based refinement, since a partition could not be computed within the time limit of one hour for most values of $k$.

Hypergraphs: it-2004, sk-2005, uk-2005

Graphs: com-Friendster, it-2004, sk-2005, soc-sinaweibo, twitter-2010, uk-2005, webbase-2001

\section{Detailed Evaluation of Coarsening Improvements}\label{app:coarsening-ablations}

\begin{figure}[h]
    \begin{minipage}{\textwidth}
    \ifpdfplots
    \includegraphics{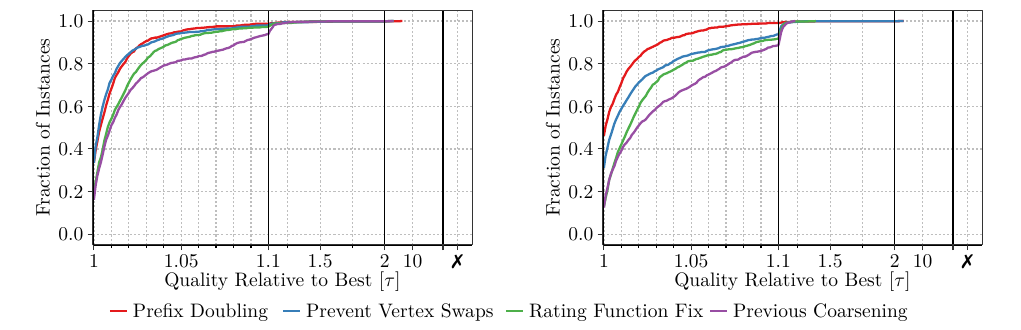}
    \else
    \tikzsetnextfilename{coarsening_ablations}%
    \input{tikz/coarsening_ablations}%
    \fi

    \end{minipage}
    \vspace{-18pt}
    \caption{Ablation study of coarsening improvements for final solution quality (left) and solution quality after initial partitioning (right).}
    \label{fig:coarsening_ablations}
\end{figure}

In Figure~\ref{fig:coarsening_ablations}, we evaluate the impact of each proposed change to the deterministic coarsening by successively enabling each component.
When considering the final solution quality, there is a noticeable improvement for both our fix of the rating function and the prevention of vertex swaps.
However, prefix doubling provides no additional improvement.
This is different when considering the quality of initial partitions instead (i.e., before uncoarsening),
which indicates that prefix doubling actually improves coarsening but the difference is compensated by refinement.

\newpage
\section{Running Time Share of Algorithmic Components for DetJet}\label{app:running-time-share}

\begin{figure}[h]
    \begin{minipage}{\textwidth}
    \ifpdfplots
    \includegraphics{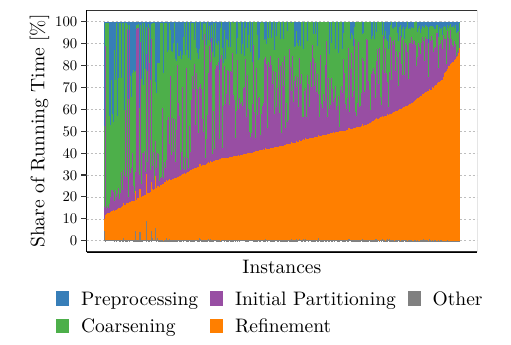}
    \else
    \tikzsetnextfilename{running_time_share_by_refinement}%
    \input{tikz/running_time_share_by_refinement}%
    \fi

    \end{minipage}
    \vspace{-10pt}
    \caption{Running time shares of different components of the DetJet configuration on all instances.
        The x-axis corresponds to the instances sorted by total refinement time.}
    \label{fig:running_time_share}
\end{figure}

Figure~\ref{fig:running_time_share} shows the relative running time of each main component of the DetJet algorithm.
We observe that Jet refinement is the most costly component, in particular for large instances.
For the smallest instances, coarsening tends to require more running time than refinement.

\end{document}